\def\MPL #1 #2 #3 {Mod.~Phys.~Lett.~{\bf#1},\  #2 (#3)}
\def\NPB #1 #2 #3 {Nucl.~Phys.~{\bf#1},\  #2 (#3)}
\def\PLB #1 #2 #3 {Phys.~Lett.~{\bf#1},\  #2 (#3)}
\def\PR #1 #2 #3 {Phys.~Rep.~{\bf#1},\ #2 (#3)}
\def\PRD #1 #2 #3 {Phys.~Rev.~{\bf#1},\  #2 (#3)}
\def\PRL #1 #2 #3 {Phys.~Rev.~Lett.~{\bf#1},\  #2 (#3)}
\def\RMP #1 #2 #3 {Rev.~Mod.~Phys.~{\bf#1},\  #2 (#3)}
\def\ZP #1 #2 #3 {Z.~Phys.~{\bf#1},\  #2 (#3)}
\def\IJMP #1 #2 #3 {Int.~J.~Mod.~Phys.~{\bf#1},\  #2 (#3)}

\def\rts{\sqrt s}
\def\eps{\epsilon}
\def\h{h}
\def\mh{m_{\h}}

\def\eg{{\it e.g.}}

\def\epem{e^+e^-}
\def\mupmum{\mu^+\mu^-}
\def\tauptaum{\tau^+\tau^-}

\def\lsim{\mathrel{\raise.3ex\hbox{$<$\kern-.75em\lower1ex\hbox{$\sim$}}}}
\def\gsim{\mathrel{\raise.3ex\hbox{$>$\kern-.75em\lower1ex\hbox{$\sim$}}}}
\def\@versim#1#2{\vcenter{\offinterlineskip
        \ialign{$\m@th#1\hfil##\hfil$\crcr#2\crcr\sim\crcr } }}

\def\ie{{\it i.e.}}

\def\gam{\gamma}

\def\anti{\overline}

\def\fbi{~{\rm fb}^{-1}}
\def\fb{~{\rm fb}}

\def\gev{\,{\rm GeV}}
\def\tev{\,{\rm TeV}}

\def\ha{A^0}

\def\tanb{\tan\beta}

\def\mt{m_t}

\def\mz{m_Z}
\def\mw{m_W}

\def\h{h}
\def\mh{m_{\h}}

\documentstyle[12pt,equations]{article}
\def\ie{{\it i.e.}}

\def\9{\phantom 0}      
\renewcommand\linebreak{\unskip\break} 
\begin{document}
\input psfig.sty
\newlength{\captsize} \let\captsize=\small 
\newlength{\captwidth}                     

%
\font\fortssbx=cmssbx10 scaled \magstep2
\hbox to \hsize{
%
%
$\vcenter{
\hbox{\fortssbx University of California - Davis}
}$
\hfill
$\vcenter{
\hbox{\bf UCD-96-14} 
\hbox{\bf IFT-10-96}
\hbox{May, 1996}
}$
}

%
\medskip
\begin{center}
\bf
DETERMINING THE TOP-ANTITOP AND ZZ COUPLINGS OF A 
NEUTRAL HIGGS BOSON OF ARBITRARY CP NATURE AT THE NLC
\\
\rm
\vskip1pc
{\bf John F. Gunion$^a$, Bohdan Grzadkowski$^{a,b}$ and Xiao-Gang He$^c$}\\
\medskip
{\em a) Davis Institute for High Energy Physics, 
University of California, Davis, CA, USA}\\
{\em b) Institute for Theoretical Physics, Warsaw University, Warsaw, Poland}\\
{\em c) School of Physics, University of Melbourne, Parkville, Australia}\\
\end{center}

\begin{abstract}
The optimal procedure for extracting the coefficients of different components 
of a cross section which takes the form of unknown coefficients times
functions of known kinematical form is developed. When applied to $\epem\to
t\anti t+$Higgs production at $\rts=1\tev$ and integrated luminosity of
$200\fbi$, we find that the $t\anti t\to$Higgs CP-even and CP-odd couplings
and, to a lesser extent, the $ZZ\to$Higgs (CP-even) coupling can be extracted
with reasonable errors, assuming the Higgs sector parameter choices yield
a significant production rate. Indeed,
the composition of a mixed-CP Higgs eigenstate
can be determined with sufficient accuracy that
a SM-like CP-even Higgs boson can be distinguished
from a purely CP-odd Higgs boson at a high level of
statistical significance, and vice versa. 

\end{abstract}

\section{Introduction}

\indent\indent 
If Higgs boson(s) (generically denoted as $\h$)
exist and are discovered at either the CERN LHC
or a future next linear $\epem$ collider (NLC), it will be extremely
important to determine both the magnitude and the CP nature of their
couplings. As reviewed in Ref.~\cite{dpfreport},
determining the CP properties of a Higgs boson through its couplings
will be especially challenging. The most promising approaches
proposed to date include: photon polarization asymmetries in $\gam\gam\to\h$
\cite{gungri}; momentum correlations among the final state $\tau$ or $t$ 
decay products appearing in $\epem\to Z\h$ and $\mupmum\to \h$ 
with $\h\to \tauptaum$ or $t\anti t$, respectively
\cite{gungrii,sonii};
and weighted cross section integrals in $pp\to t\anti t\h$ at the
LHC \cite{gunhe} and in $\epem\to t\anti t\h$ at the NLC \cite{sonietal}. The 
latter $t\anti t\h$ analyzes examined what can be accomplished using
a single observable. Ref.~\cite{gunhe} found that under ideal circumstances
a SM-like CP-even Higgs boson can
be distinguished from a purely CP-odd Higgs boson 
at a statistically significant level using $pp\to t\anti t\h$ data from 
the LHC.
Ref.~\cite{sonietal} found that a statistically significant signal 
for the CP-violating cross term
generically present in the $\epem\to t\anti t\h$ cross section for 
a Higgs boson with both CP-even and CP-odd components might be possible.
However, neither of these analyzes took full advantage of all the information
available in the cross section as a function of the kinematical variables.

In this letter, we outline the optimal technique for determining the
coefficients $c_i$ appearing in a cross section that can be
written in the generic form $d\sigma/d\phi=\sum_i c_i f_i(\phi)$,
where $\phi$ denotes the final state phase space configuration.
The application upon which we shall focus is
$\epem\to t\anti t\h$ production, where 
the $c_i$ are functions of the Higgs couplings.
By extracting the $c_i$ we can determine all the Higgs couplings and, thence,
its CP nature. This use of the full information contained
in the final state distributions, as encoded in the $c_i$, leads to 
significant improvement in the statistical precision with which the 
couplings/CP-nature
of a Higgs boson can be determined. For example, in $\epem\to t\anti t\h$
at $\rts=1\tev$, if $L=200\fbi$ and final state reconstruction efficiency
is of order $\eps=0.25$, a SM-like CP-even Higgs boson can be
distinguished from a pure CP-odd Higgs boson at roughly the $9.5\sigma$
statistical level.

\section{General Technique}

We assume that
\begin{equation}
\Sigma(\phi)\equiv {d\sigma\over d\phi}=\sum_i c_i f_i(\phi)\,,
\label{sigform}
\end{equation}
where the $f_i(\phi)$ are known functions of the location
in final state phase space, $\phi$, and the $c_i$ are
model-dependent coefficients (taken to be
dimensionless in our convention). The coefficients $c_i$ can be extracted
by using appropriate weighting functions $w_i(\phi)$ such that
$\int w_i(\phi)\Sigma(\phi)=c_i$.  In general, different
choices for the $w_i(\phi)$ are possible.  However, there is
a unique choice such that the statistical error in the determination
of the $c_i$ is minimized in the sense that the entire
covariance matrix is at a stationary point in terms of varying
the functional forms for the $w_i(\phi)$ while maintaining $\int
w_i(\phi)f_j(\phi)d\phi=\delta_{ij}$. Thus, we require
\begin{equation}
(a):~~\delta V_{ij}\propto \int \delta \left[w_i(\phi)w_j(\phi)\right]
\Sigma(\phi)d\phi=0\,,\quad
(b):~~\int \delta w_i(\phi) f_j(\phi)=0\,,
\label{variational}
\end{equation}
where $V_{ij}$ is the covariance matrix.
The weighting functions which satisfy these conditions are of the form
\begin{equation}
w_i(\phi)={\sum_j X_{ij}f_j(\phi)\over \Sigma(\phi)}\,,\quad{\rm with}~~
X_{ij}= M^{-1}_{ij}\,,\quad{\rm where}~~
M_{ik}\equiv \int {f_i(\phi)f_k(\phi)\over \Sigma(\phi)} d\phi\,,
\label{mform}
\end{equation}
since, for the $w_i(\phi)$
so defined, the constraint (b) implies the minimization
condition (a) in Eq.~(\ref{variational}).

We may then compute $c_i$ as
\begin{equation}
c_i=\sum_k X_{ik} I_k=\sum_k M_{ik}^{-1}I_k\,,\quad{\rm where}~~
I_k\equiv \int f_k(\phi) d\phi\,.
\label{ikdef}
\end{equation}
It can then be demonstrated that the covariance matrix is
\begin{equation}
V_{ij}\equiv \langle \Delta c_i\Delta c_j\rangle= 
{ M_{ij}^{-1} \sigma_T\over N}\,,
\label{cerror}
\end{equation}
where $\sigma_T=\int {d\sigma\over d\phi} d\phi$ is the integrated
cross section and $N=L_{\rm eff}\sigma_T$ is the total number of events,
with $L_{\rm eff}$ being the luminosity times efficiency.
The result of Eq.~(\ref{cerror}) applies only for the optimal
weighting functions.

We note that the above procedure is the optimal one regardless of the
relative magnitudes of the $c_i$.  Various limits of the optimal
weighting functions have previously appeared in the literature.
For example, if all of the $c_i$ are small except for $c_k$, then 
to isolate the $c_i$ ($i\neq k$) the appropriate weighting function
reduces to $w_i(\phi)\propto f_i(\phi)/f_k(\phi)$, see \eg\
Refs.~\cite{gungrii,sonii,soniearly}.

Our procedure is not altered if cuts are imposed
on the kinematical phase space over which one integrates.
Although such cuts may be required in the actual experimental analysis,
we have not included cuts in our model computations to follow.

In the $\epem\to t\anti t\h$ process, upon which we shall now focus,
in order to fully define a point in phase space
we must identify the $t$ and $\anti t$ and have no more
than one invisible particle.  Thus,
we must employ the final state mode in which one $t$ decays
leptonically and the other hadronically. Further, the $t$ and $\anti t$
must be reconstructed. The overall efficiency for the mixed leptonic-hadronic
final state decays and the double reconstruction will be denoted
by $\eps$. Then the effective luminosity is given by $L_{\rm eff}=\eps L$,
where $L$ is the total integrated luminosity. We shall take $\eps=0.25$.

If a subset, $\bar\phi$, of the kinematical
variables $\phi$ cannot be determined (as is the case for $\epem\to t\anti t\h$
if the $t,\anti t$ decay to the purely hadronic or
double-leptonic final state) the above technique
can be applied using the variables, $\hat\phi$, that {\it can} be observed
and the functions $\hat f_i(\hat\phi)\equiv \int f_i(\phi) d\bar\phi$.
This is the case even if one or more $\hat f_k$ are zero.
For example, in $\epem\to t\anti t \h$
the $f_k(\phi)$ that is a CP-odd
function of the variables $\phi$ reduces to $\hat f_k=0$
if one cannot distinguish between the $t$ and $\anti t$.

\section{Extracting Higgs Couplings in {\boldmath$\epem\to t\anti t\h$}}

We have applied the above procedure to the extraction of Higgs
couplings using the process $\epem\to t\anti t \h$.  We define
the Higgs couplings via the Feynman rules:
\begin{equation}
t\anti t\h:~~-\anti t(a+ib\gamma_5) t{g\mt\over 2\mw}\,,
\quad ZZ\h:~~c{g\mz\over \cos(\theta_W)}g_{\mu\nu}\,,
\label{coupdefs}
\end{equation}
where $g$ is the usual electroweak coupling constant.  Thus, $a$, $b$, and $c$
are defined relative to couplings of SM-magnitude. The SM Higgs boson
has $a=c=1$ and $b=0$.  A purely CP-odd Higgs boson has $a=c=0$ and $b\neq 0$;
the magnitude of $b$ depends upon the model --- we will display results
for $b=1$, which would correspond to $\tanb=1$ in a two-Higgs-doublet model
of type II (see Refs.~\cite{dpfreport,hhg} for details). This latter
choice would, in particular, apply for the CP-odd $\ha$ of the minimal
supersymmetric model (with $\tanb=1$).

The $t\anti t\h$ cross section contains five distinct terms:
$\Sigma(\phi)=\sum_{i=1}^5 c_if_i(\phi)$, where 
\begin{equation}
c_1=a^2\,;~~c_2=b^2\,;~~c_3=c^2\,;~~c_4=ac\,;~~c_5=bc\,.
\label{cdefs}
\end{equation}
Of these, the only term in $\Sigma(\phi)$ that is actually CP-violating is that
proportional to $bc$; this is the term
upon which Ref.~\cite{sonietal} focused. Our approach makes use of the 
fact that the full cross section contains additional information regarding
both $b$ and $c$.

\begin{figure}[htbp]
\let\normalsize=\captsize   
\begin{center}
\centerline{\psfig{file=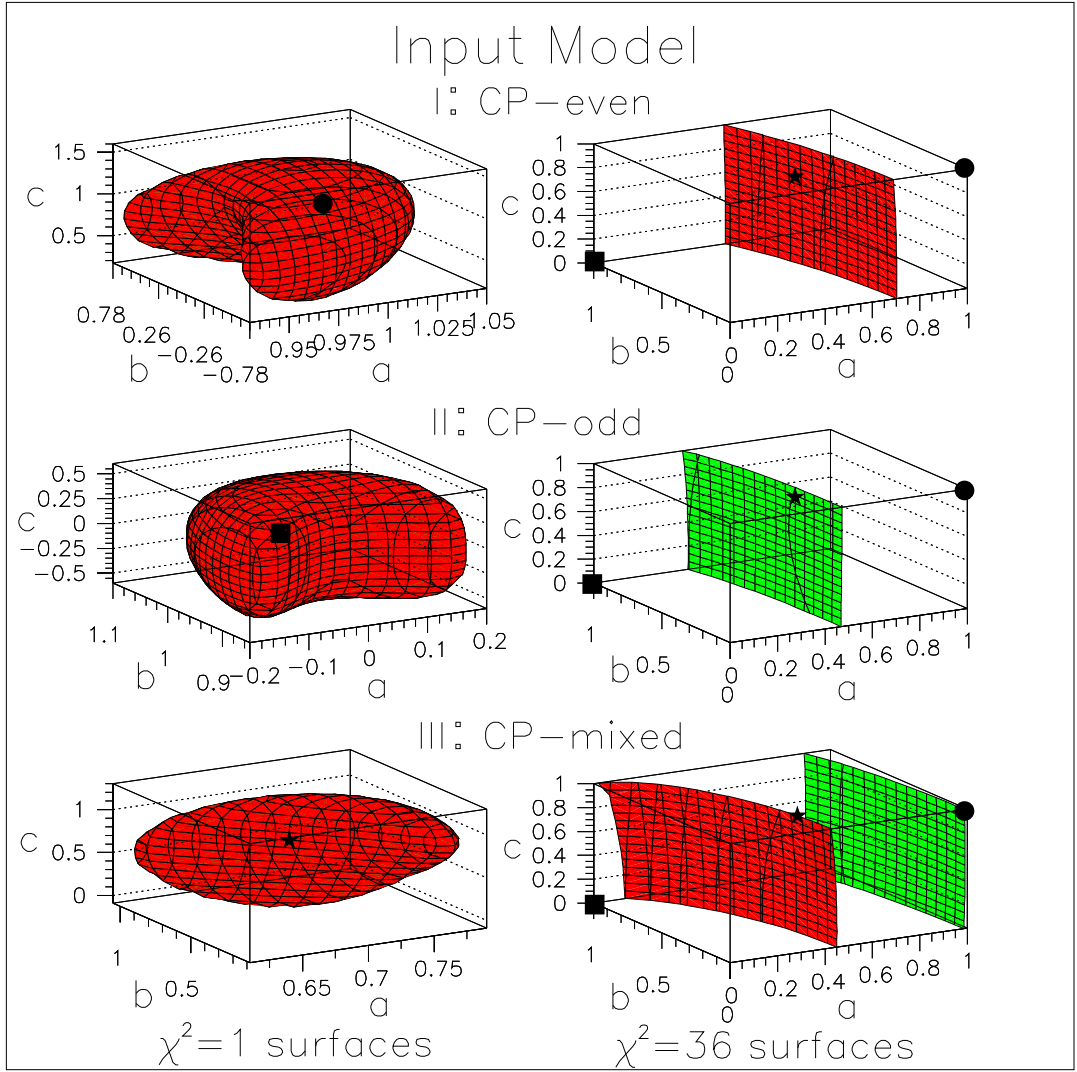,width=14.2cm}}
\begin{minipage}{12.5cm}       
\caption{Surfaces of constant $\chi^2=1$ and 36
are displayed for: I) $a=c=1$, $b=0$;
II) $a=c=0$, $b=1$; and III) $a=b=c=1/\protect\sqrt 2$.  The parameter
space locations for I, II and III are indicated by a solid bullet, square,
and star, respectively. Results are for unpolarized beams,
$\protect\rts=1\tev$, $\mh=100\gev$,
$\mt=176\gev$ and $L_{\rm eff}=50\fbi$.}
\label{contour}
\end{minipage}
\end{center}
\end{figure}

We have considered three distinct Higgs coupling cases:
\begin{itemize}
\item I) The Standard Model Higgs boson, with $a=c=1$, $b=0$.
\item II) A pure CP-odd Higgs boson, with $a=c=0$, $b=1$.
\item III) A CP-mixed Higgs boson, with $a=b=c=1/\sqrt 2$.
\end{itemize}
For unpolarized beams, $\rts=1\tev$, $\mh=100\gev$ and $\mt=176\gev$,
the integrated cross sections in cases I, II and III are
$\sigma_T=2.71$, 0.53, and $1.62\fb$, respectively.
Adopting $L_{\rm eff}=50\fbi$, we then computed
\begin{equation}
\chi^2=\sum_{i,j=1}^5 (c_i-c_i^0)(c_j-c_j^0)V^{-1}_{ij}\,, \quad {\rm with}
~~V^{-1}_{ij}={M_{ij} N\over \sigma_T}\,,
\label{chisqdef}
\end{equation}
(see Eq.~(\ref{cerror}))
as a function of location in $a,b,c$ parameter space, 
where the $c_i^0$ for a given case are computed from the
model input values of $a,b,c$ (given above) using
Eq.~(\ref{cdefs}). Surfaces of constant $\chi^2=1$
and 36 are displayed in Fig.~\ref{contour} for each of the three cases.
We have indicated
the parameter space location of models I, II and III by a solid
bullet, square, and star, respectively. 
The $\chi^2=1$ surfaces indicate the $1\sigma$ errors
on the parameter determinations.  The $\chi^2=36$ (or $6\sigma$) 
surfaces will be useful as a reference in assessing the level
at which we can distinguish the above three model cases from
one another.  

Due to the fact that the five $c_i$ are functions of only the three parameters,
$a,b,c$, the $\chi^2=1$ surfaces in Fig.~\ref{contour} are not
perfect ellipsoids. Nonetheless, we follow the usual
procedure of defining the $\pm 1\sigma$ errors in any 
one of the $a,b,c$ parameters by the largest and smallest
values that the given parameter takes as one moves about
the $\chi^2=1$ surface. (These extrema define
the locations of the two planes of constant parameter value that
are tangent to the $\chi^2=1$ surface.) 
The resulting $1\sigma$ errors are tabulated in Table~\ref{abcerrors}.
(The upper and lower limits
for $a,b,c$ employed for the $\chi^2=1$ surface plots of Fig.~\ref{contour}
are only {\it just} beyond the extrema values.)
We observe that $a$ is well determined in all cases, but especially for 
the $a\neq 0$ cases I and III. Similarly, $b$ is well determined in the 
$b\neq 0$ cases II and III. The magnitude of the error in $c$ is similar
for all three cases, and is never especially small.
Of course, a much better measurement (\eg\ $\pm5\%$ for a SM-like $\h$) of
or bound on $c$ will be available from 
inclusive $Z\h$ production; however, this does not lead to
reduced errors for $a$ and $b$.
Some improvement in the errors is possible if the electron
beam can be negatively polarized without loss
of luminosity;  the errors for $P(e^-)=-1$ are given in the table.
In what follows, we shall only consider the case of unpolarized
beams.

\def\rtw{${1\over \sqrt 2}$}
\begin{table}[hbt]
\caption[fake]{We tabulate the $1\sigma$ errors, as defined in the text, 
in $a$, $b$, and $c$
for the three Higgs coupling cases I, II and III, assuming
$\protect\rts=1\tev$, $\mh=100\gev$, $\mt=176\gev$ and $L_{\rm eff}=50\fbi$.
Results for unpolarized beams and for 100\% negative $e^-$ polarization
are given.}
\begin{center}
\begin{tabular}{|c|c|c|c|c|c|c|}
\hline
\ & \multicolumn{3}{c|}{Unpolarized $e^-$} & \multicolumn{3}{c|}{$P(e^-)=-1$} \\
Case & $a\pm\Delta a$ & $b\pm\Delta b$ & $c\pm\Delta c$ 
     & $a\pm\Delta a$ & $b\pm\Delta b$ & $c\pm\Delta c$ \\
\hline
I & 1$+0.043\atop -0.066$ & 0$+0.76\atop -0.76$ & 1$+0.51 \atop -0.82$
  & 1$+0.037\atop -0.057$ & 0$+0.70\atop -0.70$ & 1$+0.51\atop -0.77$ \\[.07in]
\hline
II & 0$+0.19\atop-0.19$ & 1$+0.093\atop-0.14$ & 0$+0.58\atop-0.58$ 
   & 0$+0.18\atop-0.18$ & 1$+0.079\atop-0.12$ & 0$+0.55\atop-0.55$ \\[.07in]
\hline
III & \rtw $+0.075\atop -0.087$ & \rtw $+0.31\atop-0.62$ & 
 \rtw $+0.57\atop-0.80$ 
    & \rtw $+0.066\atop -0.074$ & \rtw $+0.27\atop-0.46$ & 
 \rtw $+0.56\atop-0.73$ \\[.07in]
\hline
\end{tabular}
\end{center}
\label{abcerrors}
\end{table}

Most important is the ability to distinguish different Higgs
CP mixtures from one another.  Referring to Fig.~\ref{contour},
we observe the following:\footnote{We refer
to a parameter location on the $\chi^2=s^2$ surface as an $s$-sigma deviation
in the sense that the relative probability or likelihood compared
to $s=0$ is given by $\exp[-s^2/2]$, just as for a one-dimensional
parameter space. Thus, $\chi^2=36$ corresponds to relative
probability of $1.52\times 10^{-8}$. This differs from the integrated 
probability for the parameters to lie outside the $\chi^2=s^2$ surface,
which for $\chi^2=36$ is $7.49\times 10^{-8}$ for 3 parameters, \ie\ degrees
of freedom.}                                   
\begin{itemize}
\item
If the Higgs is the CP-even SM Higgs boson, then
the pure CP-odd case is well beyond even the $\chi^2=36$ surface,
and, in fact, it lies on roughly the $\chi^2\sim 90$ surface,
corresponding to discrimination at the $9.5\sigma$
statistical level.  Even the
equal CP mixture case III (the parameter location of which appears behind
the $\chi^2=36$ surface in the figure) 
is ruled out at the $4.8\sigma$ level.
\item If the Higgs is pure CP-odd, with SM $t\anti t$ coupling {\it magnitude},
then the CP-mixed and CP-even cases lie $17\sigma$ and $34\sigma$
away, respectively.
\item If the Higgs is an equal mixture of CP-even and CP-odd,
with coupling strengths specified by $a=b=c=1/\sqrt 2$, then
the SM CP-even and pure CP-odd cases I and II are both about $6.3\sigma$
away, \ie\ just a bit further away than the $\chi^2=36$ surfaces plotted.
\end{itemize}
These results improve if the $e^-$ beam has negative polarization.
The discrimination abilities are summarized in Table~\ref{modeldisc}.

\begin{table}[hbt]
\caption[fake]{We tabulate the number of standard deviations, $\sqrt{\chi^2}$,
at which a given input model (I, II or III) can be distinguished from
the other two models, assuming
$\protect\rts=1\tev$, $\mh=100\gev$, $\mt=176\gev$ and $L_{\rm eff}=50\fbi$.
Results for unpolarized beams and for 100\% negative $e^-$ polarization
are given.}
\begin{center}
\begin{tabular}{|c|c|c|c|c|c|c|}
\hline
\ & \multicolumn{3}{c|}{Unpolarized $e^-$} & \multicolumn{3}{c|}{$P(e^-)=-1$} \\
\ & \multicolumn{3}{c|}{Trial Model} & \multicolumn{3}{c|}{Trial Model} \\
Input Model &  I & II & III & I & II & III \\
\hline
I   &  -  & 9.5 & 4.8  &  -  & 11  & 5.5 \\
II  & 34  &  -  &  17  & 40  & -   & 20  \\
III & 6.3 & 6.3 &  -   & 7.3 & 7.3 &  -  \\
\hline
\end{tabular}
\end{center}
\label{modeldisc}
\end{table}
 
We can also analyze our ability to
determine that the CP-violating component of $\Sigma(\phi)$,
proportional to $c_5\equiv bc$, is non-zero. 
We consider model III (the only one of our three models for which $bc\neq 0$).
We plot the $\chi^2=1$ ($1\sigma$) surface in $a$, $b$, and $bc$ space
and look for the extrema of $bc$.  We find that these extrema
occur for $a\sim b\sim 1/\sqrt{2}$ and that $bc$ can range from
$-.05$ to $+.91$, assuming $L_{\rm eff}=50\fbi$ and unpolarized beams.
Clearly, we are not far from establishing
a non-zero signal at the $1\sigma$ level.
For twice as much luminosity, $L_{\rm eff}\sim 100\fbi$,
the extrema of $bc$ on the $1\sigma$ surface
are +.15 and +.79, and a non-zero value
of $bc$ would have been established at better than the $1\sigma$ level.
At the $1\sigma$ level, 
$L_{\rm eff}=50\fbi$ upper bounds on $|c_5|=|bc|$
in models I and II are 0.65 and 0.55, respectively.
The above results are all somewhat better than obtained
for these same models 
using either of the observables (${\cal O}$ or ${\cal O}_{\rm ropt}$)
employed in Ref.~\cite{sonietal}.

\section{Final Remarks and Conclusions}

\indent\indent
In this letter, we have outlined the optimal technique for extracting
the coefficients that appear in a general cross section which is 
a sum of model-dependent coefficients times known kinematical functions.
Application of this technique to $\epem\to t\anti t\h$ results in
good prospects for pinning down the CP nature of
the $\h$ at a $1\tev$ $\epem$ collider operating at an expected luminosity
of $L=200\fbi$, provided only that the $\h$ has a reasonable production
cross section (roughly $\gsim 0.5\fb$) and that the $t\anti t\h$
final state can be reconstructed with reasonable efficiency
(roughly $\eps \gsim 0.2$). In particular, for a Higgs
mass of 100 GeV and unpolarized beams,
it will be possible to demonstrate, using $\epem\to t\anti t\h$
production only, that a SM Higgs boson has the expected 
CP-even couplings to $t\anti t$ and $ZZ$ within $+ 4.3\%\atop -6.6\%$ 
and $+51\%\atop -82\%$, respectively, and that
any CP-odd coupling to $t\anti t$ is less than 76\% of the CP-even
SM strength. The precision with
which both the CP-odd and CP-even 
$t\anti t$ Higgs couplings can be determined is somewhat
improved for a negatively polarized electron beam, assuming
there is no loss of luminosity.
Most importantly, the coefficients of the various terms in the
$\epem\to t\anti t\h$ cross section can be determined well enough
that Higgs CP mixtures that are significantly different
from one another can generally be distinguished
at a substantial (sometimes very substantial)
level of statistical significance. 

We have implicitly assumed that the systematic error in the overall
normalization of the $t\anti t\h$ cross section will be 
relatively small, \eg\ $\lsim \pm 5\%$.  If this is not the
case, then one can focus on the ratios of the different cross section
coefficients
to one another. Our technique is easily adapted to this situation.

\section{Acknowledgements}

This work was supported in part by Department of Energy under
grant No. DE-FG03-91ER40674,
by the Davis Institute for High Energy Physics, by the
Committee for Scientific Research (Poland), and by Maria Sklodowska-Curie
Joint Fund II (Poland-USA) under grant No. MEN/NSF-96-252.
XGH was supported in part by the Australian Research Council. BG and XGH
would like to thank the Davis Institute for High Energy Physics for
hospitality. We would like to thank I. Fissiak and R. Nowak for 
comments on our statistical analysis.

\clearpage
 
\end{document}